\documentclass[aps,prc,preprint,amsmath,amssymb,superscriptaddress]{revtex4}
\usepackage{graphicx}  % needed for figures
\usepackage{dcolumn}   % needed for some tables
\usepackage{bm}        % for math
\usepackage{amssymb}   % for math
\usepackage{enumerate}
\usepackage{amsmath}
\usepackage{tabularx}
\usepackage{mathtools}
\usepackage{xcolor}
\usepackage{ctable}
\usepackage{rotating}
\usepackage{wrapfig}

\begin{document}

% The following information is for internal review, please remove them for submission
\widetext
%\leftline{Version xx as of \today}
%\leftline{Primary authors: Joe E. Physics}
%\leftline{To be submitted to PRL}
%\leftline{Comment to {\tt d0-run2eb-nnn@fnal.gov} by xxx, yyy}
%\centerline{\em D\O\ INTERNAL DOCUMENT -- NOT FOR PUBLIC DISTRIBUTION}
% the following line is for submission, including submission to the arXiv!!
%\hspace{5.2in} \mbox{Fermilab-Pub-04/xxx-E}

\title{Quest for consistent modelling of statistical decay of the compound nucleus}
%\input author_list.tex       % D0 authors (remove the first 3 lines
                             % of this file prior to submission, they
                             % contain a time stamp for the authorlist)
                             % (includes institutions and visitors)
\author{Tathagata Banerjee}
\email{he.tatha@gmail.com}
\affiliation{Nuclear Physics Group, Inter University Accelerator Centre,
Aruna Asaf Ali Marg, Post Box 10502, New Delhi 110067, India}
\author{S. Nath}
\affiliation{Nuclear Physics Group, Inter University Accelerator Centre,
Aruna Asaf Ali Marg, Post Box 10502, New Delhi 110067, India}
%\author{Jhilam Sadhukhan}
%\affiliation{Physics Group, Variable Energy Cyclotron Centre, 
%1/AF Bidhan Nagar, Kolkata 700 064, India}
\author{Santanu Pal}
\affiliation{Nuclear Physics Group, Inter University Accelerator Centre,
Aruna Asaf Ali Marg, Post Box 10502, New Delhi 110067, India}

\date{\today}

\begin{abstract}
 A statistical model description of heavy ion induced fusion-fission reactions is
presented where shell effects, collective enhancement of level density, tilting away
effect of compound nuclear spin and dissipation are included. It is shown that the
inclusion of all these effects provides a consistent picture of fission where fission
hindrance is required to explain the experimental values of both pre-scission neutron
multiplicities and evaporation residue cross-sections in contrast to some of the
earlier works where a fission hindrance is required for pre-scission neutrons
but a fission enhancement for evaporation residue cross-sections.

\end{abstract}

%\pacs{}
\maketitle

%\section{\label{sec:level1}First-level heading}
% sections are not used for PRL papers

\section{Introduction}

Heavy ions impinging upon a heavy nucleus at beam energies below and above the
Coulomb barrier had led to a number of important discoveries, both in nuclear
structure as well as in reactions. While disappearance of pairing correlation
with increasing spin of a compound nucleus (CN) is observed in `backbending'
phenomenon in nuclear structure \cite{Stephens1975}, a diffusion mechanism of
momentum and mass between the projectile and target is found in the strongly
damped collisions \cite{Schroder1977} and a fission hindrance in fusion-fission
reactions \cite{Hinde1986, Newton1988}. Fission hindrance manifests in
observation of excess neutrons from the CN before fission with respect to the
standard statistical model predictions, indicating a slowing down of fission
process in comparison to the fission rate from the transition state theory as
given by Bohr and Wheeler \cite{BW}. Fission hindrance is found to decrease
with increase of mass and fissility of the CN formed in heavy ion induced
fusion-fission reactions \cite{Thonnensen}. It was early recognized that a
dissipative fission dynamics can account for the fission hindrance and the
resulting delayed onset of fission \cite{Hinde1986}.

Fission hindrance is expected to impact not only the multiplicity of
pre-scission neutrons and other light evaporated particles or photons but also
the fission and evaporation residue (ER) cross-sections of fusion-fission 
reactions. It is observed that at least two or more input parameters in
statistical model (SM) calculations, namely those defining the fission barrier,
the level density parameter, the fission delay time and the dissipation 
coefficient are required to be adjusted for simultaneous fitting of both the
fission/ER and the pre-scission neutron multiplicity excitation functions
\cite{Hinde1986, Newton1988, LestonePRC79, MahataPRC74, MahataPRC92}. Different
values of parameter sets are found necessary for different systems. Further, an
increase of fission probability is usually required by way of reducing the
fission barrier while fitting fission/ER cross sections in standard statistical
model calculations \cite{SagaidakPRC79,VickyPRC89} whereas a slowing down of
fission is demanded to reproduce experimental pre-scission neutron
multiplicities \cite{Hinde1986}. This apparent contradiction clearly points to
an inadequate modelling of compound nucleus decay. 
    
In the present work, we show that a consistent description of both the
pre-scission neutron multiplicity and fission/ER cross sections can be obtained
with only one adjustable parameter when all the other factors which influence
the fission and various evaporation widths are taken into account. To this end,
we consider the shell correction effects on both the nuclear level density and
the fission barrier, the collective enhancement of level density (CELD) and the
effect of orientation ($K$ state) degree of freedom of CN spin on fission
width. We also include the effect of nuclear dissipation on fission width.
Different combinations of the above effects have been considered by many
authors in the past for statistical model analysis of fusion-fission reactions
\cite{Tatha2015,Tatha2016,LestonePRC79,MayorovPRC92}. All the four effects have
been included in the statistical model analysis of pre-scission and
post-scission multiplicity by Yanez et al. \cite{YanezPRL112}. Here we consider
all the effects for statistical model analysis of both pre-scission neutron
multiplicity and fission/ER cross sections.

\section{Description of Model and Results}

The various input quantities for the statistical model calculation are chosen
as follows. We argue that nuclear properties which are well defined and well
understood from independent studies should be used without any further
modification in the statistical model of compound nuclear decay. Accordingly,
we first obtain the macroscopic part of the fission barrier from the
finite-range liquid drop model (FRLDM) which fits the systematic behaviour of
ground state masses and the fission barriers at low angular momentum $\ell$ for
nuclei over a wide mass range \cite{SierkPRC33}. The full fission barrier
$B_{\textrm{f}}(\ell)$ of a nucleus carrying angular momentum is then obtained
by incorporating shell correction to the FRLDM barrier \cite{MahataPRC92}. The
shell correction term $\delta$ is given as the difference between the
experimental and the liquid-drop model (LDM) masses
$\left(\delta = M_{\textrm{exp}} - M_{\textrm{LDM}}\right)$. The fission
barrier then is given as 

\begin{equation}
\label{barrier}
B_{\textrm{f}}(\ell) = B_{\textrm{f}}^{\textrm{LDM}}(\ell) -
\left( \delta_{\textrm{g}} - \delta_{\textrm{s}} \right)
\end{equation}

\noindent
where, $B_{\textrm{f}}^{\textrm{LDM}}(\ell)$ is the angular momentum dependent
FRLDM fission barrier and $\delta_{\textrm{g}}$ and $\delta_{\textrm{s}}$ are
the shell correction energies for ground-state and saddle configurations,
respectively. The ground-state shell corrections are taken from Ref.
\cite{ Myers1994}. For $\delta_{\textrm{g}}$ and $\delta_{\textrm{s}}$, the
prescription given in Ref. \cite{MyersNP81} for deformation dependence of shell
correction is used which yields a very small value of shell correction at large
deformations and full shell correction at zero deformation. 

We next consider the level density parameter `$a$', which is taken from the
work of Ignatyuk et al. \cite{Ignatyuk1975}, who proposed the following form
which includes shell effects at low excitation energies and goes over to its
asymptotic form at high excitation energies 

\begin{equation}
\label{level}
a(E^{*}) = \tilde{a} \left[1+\frac{g(E^{*})}{E^{*}}\delta \right] ,
\end{equation}

\noindent
where, $g(E^{*}) = 1 - \exp\left(-\frac{E^{*}}{E_{\textrm{D}}}\right)$,
$\tilde{a}$ is the asymptotic level density and $E_{\textrm{D}}$ is a parameter
which decides the rate at which the shell effects disappear with an increase in
the intrinsic excitation energy ($E^{*}$). A value of 18.5 MeV is used for
$E_{\textrm{D}}$, which was obtained from an analysis of $s$-wave neutron
resonances \cite{Reisdorf1981}. The shape-dependent asymptotic level density 
is also taken from Ref.\cite{Reisdorf1981}. 

\begin{figure*}[ht!]
\begin{center}
\includegraphics[width=\textwidth,trim=0.0cm 0.0cm 0.0cm 0.0cm,clip]{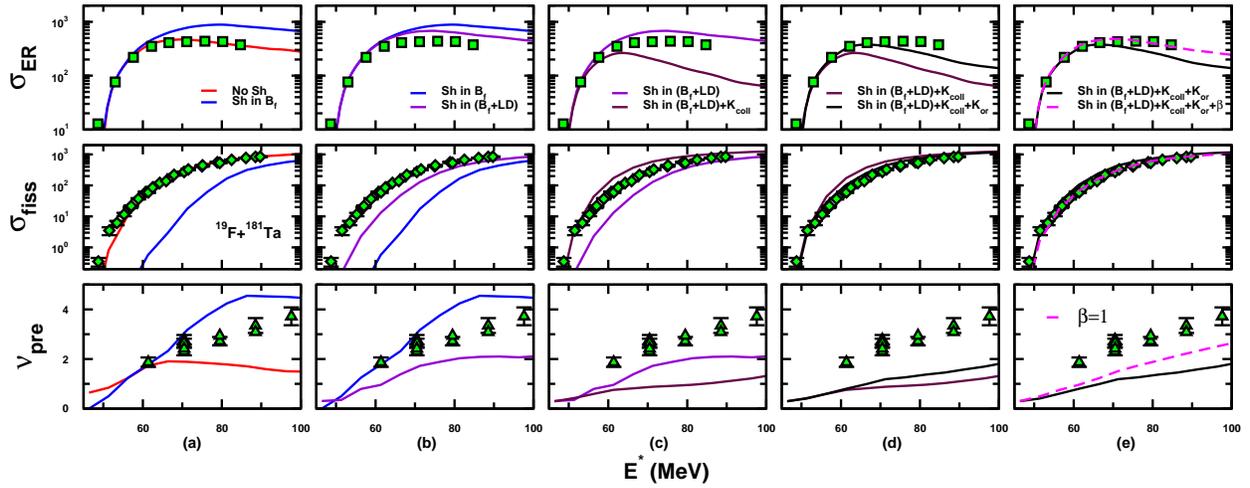}
\caption{\label{FIG1} Comparison of measured $\sigma_{\textrm{ER}}$,
$\sigma_{\textrm{fiss}}$ and $\nu_{\textrm{pre}}$ with SM predictions for the
reaction $^{19}$F+$^{181}$Ta \cite{Hinde1986, Charity1986, Newton1988}. Panel
(a): Red lines denote the results with no shell effects and blue lines
represent results with shell correction in fission barrier ($B_{\textrm{f}}$)
only; panel (b): blue lines same as in panel (a) and violet lines are for shell
correction in both $B_{\textrm{f}}$ and ground state level density (LD); panel
(c): violet lines same as in panel (b) and maroon lines stand for results from
shell corrected $B_{\textrm{f}}$ and LD plus CELD; panel (d): maroon lines same
as in panel (c) and black lines indicate shell corrected $B_{\textrm{f}}$ and
LD, inclusion of CELD and $K$-orientation effects; panel (e): black lines same
as in panel (d) and dashed magenta lines include all the aforementioned effects
and a suitable dissipation.}
\end{center}
\end{figure*}

The effect of collective motion in nuclear Hamiltonian on nuclear level density
was investigated earlier by Bj{\o}rnholm, Bohr and Mottelson \cite{BBM1974}.
They observed that the total level density $\rho (E^{*})$ can be obtained from
the intrinsic level density $\rho_{\textrm{intr}} (E^{*})$ as 

\begin{equation}
\rho (E^{*}) = K_{\textrm{coll}}(E^{*}) \rho_{\textrm{intr}} (E^{*})
\end{equation}

\noindent
where, $K_{\textrm{coll}}$ is the collective enhance factor. A significant role
of collective enhancement of level density (CELD) in reproducing the mass
distribution from fragmentation of heavy nuclei was observed by Junghans
\textit{et al.} \cite{JunghansNPA629}. CELD is also found to be important for
calculating the survival probability of super-heavy nuclei
\cite{MayorovPRC92,YanezPRL112}. Evidence of CELD has also been found in the
spectrum of evaporated neutrons from deformed compound nuclei
\cite{PratapPRC88}. 

The effect of CELD is incorporated in the present calculation following the
work of Zagrebaev \textit{et al.} \cite{ZagrebaevPRC65} where a smooth
transition from a vibrational enhancement $K_{\textrm{vib}}$ to a rotational
enhancement $K_{\textrm{rot}}$ for a nucleus with quadrupole deformation
$|\beta_{2}|$ was achieved through a function $\varphi$($|\beta_{2}|$) as follows 

\begin{equation}
\label{kcol}
K_{\textrm{coll}}(|\beta_{2}|) = [K_{\textrm{rot}}\varphi(|\beta_{2}|)+K_{\textrm{vib}}(1-\varphi(|\beta_{2}|))]\ f(E^{*})\\
\end{equation}
where,
\begin{equation}
\varphi(|\beta_{2}|) = \left[1+\exp\left(\frac{\beta_{2}^{0}-|\beta_{2}|}{\Delta\beta_{2}}\right)\right]^{-1}.
\end{equation}

\noindent
The values of $\beta_{2}^{0}= 0.15$ and $\Delta\beta_{2}= 0.04$ are taken from
Ref. \cite{Ohta2001}. The Fermi function $f(E^{\textrm{*}})$ accounts for the
damping of collectivity with increasing excitation enegy $E^{\textrm{*}}$ and
is given as 

\begin{equation}
f(E^{\textrm{*}})= \left[1+\exp\left(\frac{E^{*}-E_{\textrm{cr}}}{\Delta E}\right)\right]^{-1}
\end{equation}

\noindent
with $E_{\textrm{cr}}=40$ MeV and $\Delta E= 10$ MeV \cite{ JunghansNPA629}.
The rotational and vibrational enhancement factors are taken as
$K_{\textrm{rot}} = \frac{\tau_{\perp}T}{\hbar^{2}}$, and 
$K_{\textrm{vib}} = \textit{e}^{0.055 \times A^{\frac{2}{3}}\times T^{\frac{4}{3}}}$,
where $A$ is the nuclear mass number, $T$ is the nuclear temperature and
$\tau_{\perp}$ is the rigid body moment of inertia perpendicular to the
symmetry axis \cite{Ignatyuk1985}. 

We next include the effect of $K$-degree (angular momentum component of the CN
along symmetry axis) of freedom in fission width. The angular momentum of a CN
can change its orientation from its initial direction along the perpendicular
to the symmetry axis ($K=0$) to non-zero values of $K$ due to the coupling of
the $K$-degree of freedom with intrinsic nuclear motion \cite{LestonePRC79}.
Assuming a fast equilibration of the $K$-degree of freedom, the fission width
can be expressed as 
\cite{LestonePRC59}  

\begin{equation}
\label{KOR}
\Gamma_{\textrm{f}} (E^{*}, \ell) = \Gamma_{\textrm{f}}(E^{*},\ell, K=0) \frac{\left(K_{0}\sqrt{2\pi}\right)}{2\ell+1}\ \textrm{erf} 
\left(\frac{\ell+\frac{1}{2}}{K_{0}\sqrt{2}}\right)
\end{equation}

\noindent
with $K_{0}^{2} = \frac{\tau_{\textrm{eff}}}{\hbar^{2}} T_{\textrm{sad}}$,
where, $\tau_{eff}$ is the effective moment of inertia
$\left(\frac{1}{\tau_{eff}} = \frac{1}{\tau_{\parallel}} -
\frac{1}{\tau_{\perp}}\right)$, $\tau_{\perp}$ and $\tau_{\parallel}$ being 
the moments of inertia at saddle of the nucleus perpendicular to and about the
nuclear symmetry axis and erf(x) is the error function.

Using  the various quantities as given in the above, statistical model
calculations are performed for pre-scission neutron multiplicity, fission and
ER cross sections for a number of systems. The fission width
$\Gamma_{\textrm{f}} (E^{*}, \ell, K=0)$ in Eq. \ref{KOR} is obtained from the
transition state theory due to Bohr and Wheeler \cite{BW} where shell-corrected
fission barrier (Eq. \ref{barrier}) and level density parameter
(Eq. \ref{level}) are used. CELD (Eq. \ref{kcol}) is also included in the level
densities at both the ground state and at the saddle in fission width
calculation. CELD at ground state is calculated using experimental values of
$|\beta_{2}|$ for deformed nuclei. The particle and $\gamma$ emission widths are
obtained from the Weisskopf formula as given in Ref. \cite{FrobrichPR}. Shell
correction and CELD are applied to the level densities of both the parent and
the daughter nuclei. 

In the present SM calculation, a CN is followed in time over small time steps
and at each time step the fate of the compound nucleus is decided by a Monte
Carlo sampling using the particle, $\gamma$ and fission widths
\cite{GargiPRC65}. The process continues till fission occurs or an ER is
formed. Fission in SM corresponds to crossing of the saddle point deformation
by the CN. During transition from the saddle to the scission, the CN can emit
further neutrons which would contribute to the pre-scission multiplicity. The
saddle-to-scission time interval $\tau_{ss}^{0}$ is taken from Ref.
\cite{NixNPA130} and the number of neutrons emitted during this period is
calculated using the neutron decay width.

The systems chosen for the present study are
(a) $^{16}$O+$^{154}$Sm \cite{Leigh1995,Zebelman1973,Gavron1981}, 
(b) $^{19}$F+$^{169}$Tm \cite{Charity1986,Newton1988},
(c) $^{16}$O+$^{184}$W \cite{Leigh1988,Shidling2006,Bemis1986,Forster1987,Hinde1992}, 
(d) $^{19}$F+$^{181}$Ta \cite{Hinde1986,Charity1986,Newton1988},
(e) $^{18}$O+$^{192}$Os \cite{Hinde1986,Charity1986,Newton1988}, 
(f) $^{16}$O+$^{197}$Au \cite{Brinkmann1994,Sikkeland1964,Newton1988} and
(g) $^{16}$O+$^{208}$Pb \cite{Brinkmann1994,Rossner1992,Morton1995,Back1985}.
They cover a broad range of fissility (0.6 to 0.763) of the compound nuclei. 
Further, non-compound nuclear fission processes are expected to be small for
the above highly asymmetric systems \cite{ShamlathPRC95}. It may therefore be
assumed that the entire incident flux leads to CN formation and consequently
the SM becomes applicable to calculate various observables in the above
reactions. 

In order to illustrate the effects of shell, CELD and the $K$-degree of
freedom, results for one reaction, namely $^{19}$F+$^{181}$Ta forming the CN
$^{200}$Pb, are presented in Fig. \ref{FIG1}. 
We first calculate pre-scission
neutron multiplicity, fission and ER excitation functions without considering
any shell correction, CELD or $K$-degree of freedom. Shell correction to
fission barrier (Eq. \ref{barrier}) only is added in the next calculation. Both
the results are given in Fig. \ref{FIG1} 
 (a) along with the experimental data.
The observation that addition of shell correction to fission barrier increases
$\nu_{\textrm{pre}}$ and decreases $\sigma_{\textrm{fiss}}$ (and
correspondingly increases $\sigma_{\textrm{ER}}$) for the present system can be
easily understood from the following relation \cite{Hinde1986} 

\begin{equation}
\frac{\Gamma_{\textrm{f}}}{\Gamma_{\textrm{n}}} \approx \frac{\textit{e}^{2\sqrt{a_{\textrm{s}}(E^{*}-B_{\textrm{f}})}}}
{\textit{e}^{2\sqrt{a_{\textrm{g}}(E^{*}-B_{\textrm{n}})}}}
\end{equation}

\noindent
where, $\Gamma_{\textrm{n}}$ and $B_{\textrm{n}}$ denote the neutron width and
binding energy, respectively. $a_{\textrm{s}}$ and $a_{\textrm{g}}$ represent
the level density parameters at the saddle and the ground state configurations,
respectively. Shell corrections for the CN $^{200}$Pb and other nuclei
populated by light particle evaporation are negative quantities and hence cause
increase in the respective fission barriers (Eq. \ref{barrier}) resulting in
decrease of the fission widths and consequently reduction in the fission
cross-sections.

Shell correction is next added to the level density parameter and Fig. \ref{FIG1}
(b) shows the results. Shell correction essentially affects the
level density at the ground state and on account of it being a negative
quantity reduces $a_{g}$ (Eq. \ref{level}) and thereby increases the fission
probability and hence the fission cross-sections. Consequently,
$\sigma_{\textrm{ER}}$ and $\nu_{\textrm{pre}}$ decreases as we see in Fig. \ref{FIG1} 
(b).

We now include CELD in the level densities for both the initial and final
states in calculation of particle and $\gamma$ decay widths. CELD is also
included in the level densities at the ground state and the saddle
configuration to obtain the fission widths. The saddle shape being highly
deformed, CELD at saddle is of rotational type while it is of vibrational
nature for spherical nuclei at ground state. The typical value of
$K_{\textrm{vib}}$ is $\sim$1 \textendash 10. $K_{\textrm{rot}}$ takes the
value $\sim$100 \textendash 150. By definition of CELD, the lower limit of Kcoll 
is set as unity. Since, the transition-state fission width is
determined by the ratio of the number of levels available at the saddle to
those at the ground state, CELD can substantially increase the fission width
for spherical nuclei. This is reflected in SM results given in Fig. \ref{FIG1}
 (c) for the present system for which the compound nuclei formed at various
stages of evaporation are spherical at the ground state and thus an enhancement
of fission cross-section is observed. It may however be remarked that for
nuclei with strong ground state deformation, the enhancement factors at both
the saddle and the ground state are of rotational type with similar magnitudes
and this would result in a marginal effect on the fission width.

The $K$-degree of freedom is next included in the SM calculation through its
effect on the fission width (Eq. \ref{KOR}). The tilting of the angular
momentum vector away from the normal direction to the symmetry axis increases
the angular momentum dependent fission barrier and reduces the fission width
\cite{LestonePRC59}. This results in a decrease in $\sigma_{\textrm{fiss}}$ and
increase of $\sigma_{\textrm{ER}}$ and $\nu_{\textrm{pre}}$ as shown in Fig. \ref{FIG1} 
(d).

\begin{figure*}[ht!]
\begin{center}
\includegraphics[width=\textwidth,trim=0.0cm 0.0cm 0.0cm 0.0cm,clip]{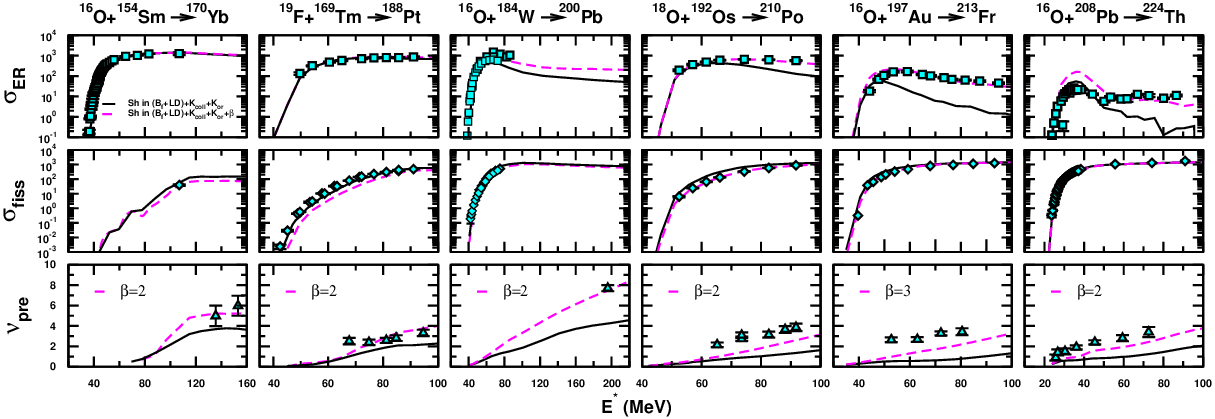}
\caption{\label{FIG2} Measured and calculated $\sigma_{\textrm{ER}}$,
$\sigma_{\textrm{fiss}}$ and $\nu_{\textrm{pre}}$ for
  (a) $^{16}$O+$^{154}$Sm \cite{Leigh1995,Zebelman1973,Gavron1981},
  (b) $^{19}$F+$^{169}$Tm \cite{Charity1986,Newton1988},
  (c) $^{16}$O+$^{184}$W  \cite{Leigh1988,Shidling2006,Forster1987,Hinde1992},
  (d) $^{18}$O+$^{192}$Os \cite{Hinde1986,Charity1986,Newton1988},
  (e) $^{16}$O+$^{197}$Au \cite{Brinkmann1994,Sikkeland1964,Newton1988} and
  (f) $^{16}$O+$^{208}$Pb \cite{Brinkmann1994,Rossner1992,Morton1995,Back1985}.
Legends are same as in Fig. \ref{FIG1} (e). See text for details.}
\end{center}
\end{figure*}

We thus find that both $\nu_{\textrm{pre}}$ and $\sigma_{\textrm{ER}}$ are
underestimated and $\sigma_{\textrm{fiss}}$ is overestimated when all the
factors which can influence the widths of various decay channels including
fission are taken into account. This immediately suggests that fission
hindrance is required to reproduce both $\sigma_{\textrm{ER}}$ and
$\nu_{\textrm{pre}}$ (and also $\sigma_{\textrm{fiss}}$). In the framework of a
dissipative dynamical model of fission, a reduction in fission width can be
obtained from the Kramers-modified fission width given as \cite{Kramers}

\begin{equation}
\Gamma_{\textrm{K}} = \Gamma_{\textrm{f}} \left\{ \sqrt{1+\left(\frac{\beta}{2\omega_{\textrm{s}}}\right)^{2}}-\frac{\beta}{2\omega_{\textrm{s}}} \right\}
\end{equation}

\noindent 
where, $\Gamma_{\textrm{f}}$ is the Bohr-Wheeler fission width obtained with 
shell corrected level densities, CELD and K-orientation effect, $\beta$ is the
reduced dissipation coefficient (ratio of dissipation coefficient to inertia)
and $\omega_{\textrm{s}}$ is the local frequency of a harmonic oscillator
potential which osculates the nuclear potential at the saddle configuration and
depends on the spin of the CN \cite{JhilamPRC78}. 

The main mechanism of energy dissipation in bulk nuclear dynamics at low
excitation energies ($T \sim$ a few MeV) is expected to be of one-body type
which arises due to collisions of the nucleons with the moving mean-field
\cite{Blocki1978, Wada1993}. The precise nature of nuclear one-body dissipation
is yet to be established though a shape-dependence \cite{PalPRC57} and 
temperature-dependence \cite{HofmannPRC64} of one-body dissipation coefficient
have been suggested on theoretical grounds. The values of $\beta$ obtained from
fitting experimental data vary in the range
$\sim$(1 \textendash 20)$\times 10^{21}$ s$^{-1}$
\cite{VickyPRC86, RohitPRC87, DioszegiPRC61}. Thus $\beta$ is the least
precisely determined input parameter among all the others and hence is treated
as the only adjustable parameter in the present SM calculation. 

Apart from the fission width, the saddle-to-scission time interval also changes
with inclusion of dissipation and is given as \cite{HofmannPLB122} 

\begin{equation}
\tau_{\textrm{ss}} = \tau_{\textrm{ss}}^{0} \left\{\sqrt{1+\left(\frac{\beta}{2\omega_{\textrm{s}}}\right)^{2}}+\frac{\beta}{2\omega_{\textrm{s}}}\right\} .
\end{equation}
    
Further, the fission width reaches its stationary value in a dissipative
dynamics of fission after the elapse of a build up or transient time
$\tau_{\textrm{f}}$ and the dynamical fission width is given as
\cite{BhattPRC33}

\begin{equation}
\Gamma_{\textrm{f}}(t) = \Gamma_{\textrm{K}} \left\{1-\textit{e}^{-\frac{2.3t}{\tau_{\textrm{f}}}}\right\}
\end{equation}

\noindent
which is used in the time evolution of the system in the present calculation.
The neutron multiplicities and fission / ER cross-sections calculated with
$\beta = 1 \times 10^{21}$ s$^{-1}$ are given in Fig. \ref{FIG1} 
 (e) which fit
the fission / ER excitation functions reasonably well and underestimate the
neutron multiplicities to some extent.

Fig. \ref{FIG2}, shows the SM predictions for the other six systems where all
the effects, namely shell, CELD and $K$-orientation are included (continuous black lines)
along with the experimental data. Neutron multiplicity $\nu_{\textrm{pre}}$ is
found to be underestimated for all the systems. The $\sigma_{\textrm{ER}}$ are
also underestimated (consequently $\sigma_{\textrm{fiss}}$ overestimated) for
the systems with  compound nuclear masses $A_{\textrm{CN}} \geq 200$ whereas
for the two systems with $A_{\textrm{CN}} < 200$, the calculated ER and fission
excitation functions are very close to the experimental data. SM results with
fission hindrance are also given in Fig. \ref{FIG2} (dashed magenta lines) where the
value of $\beta$ is adjusted to reproduce the experimental ER excitation
functions for $A_{\textrm{CN}} \geq 200$ and $\nu_{\textrm{pre}}$ for
$A_{\textrm{CN}} < 200$.

\section{Discussions}

We first note in Fig. \ref{FIG2} that $\nu_{\textrm{pre}}$ is underestimated
for $\beta$ values which reproduce the ER (and fission) excitation functions
for $A_{\textrm{CN}} \geq 200$ systems. Therefore, these values of $\beta$
though are adequate for pre-saddle fission dynamics are not large enough to
cause sufficient delay for emitting large number of neutrons in the 
saddle-to-scission evolution of the CN. It has also been observed earlier that
a shape-dependent $\beta$ with larger values in the post-saddle region is
required in order to explain the experimental neutron multiplicities for heavy
systems \cite{ DioszegiPRC61, FrobrichNPA556}. It may be pointed out
that though dissipative dynamical models such as the Langevin equation is
better suited to describe fission dynamics with shape-dependent dissipation
\cite{JhilamPRC81}, the fission cross-sections are underestimated by the
dynamical model possibly due to non-inclusion of CELD
\cite{GargiPRC65, NadtochyPRC85}. It may however be noted that a hybrid 
approach, in which a statistical model with CELD provides the flux across the
saddle point coupled with a Langevin dynamical calculation in the
saddle-to-scission region, could provide a better description for heavy systems
\cite{Vanin1999}. The present study indicates a pre-saddle dissipation strength
of $\beta = $ (1 \textendash 3) $\times 10^{21}$ s$^{-1}$ over a broad range of
excitation energies for $A_{\textrm{CN}} \geq 200$ nuclei.    

For the lighter systems ($A_{\textrm{CN}} < 200$), it is observed that though
experimental values of $\nu_{\textrm{pre}}$ can be reproduced with
$\beta = 2 \times 10^{21}$ s$^{-1}$, the fission cross-sections are
underestimated for the CN $^{188}$Pt. The former observation corresponds to the
fact that the saddle configuration is more deformed for lighter nuclei than
heavier ones and this makes the saddle-to-scission transition of shorter
duration and consequently less number of saddle-to-scission neutrons for
lighter compound nuclei. Thus, pre-saddle neutrons seem to account for the
experimental numbers. The underestimation of $\sigma_{\textrm{fiss}}$ 
for $^{188}$Pt may possibly be traced to its ground state deformation
($\beta_{2} = 0.18$). We have pointed out earlier that CELD effect on fission
channel is weaker for CN with ground state deformation than spherical nuclei.
We assume in the present work that ground state deformation persists at all
excitations. However, it is possible that the nucleus tends to become spherical
with increasing excitation energy \cite{Goodman1986}. The consequence of this
aspect in SM calculation requires further investigations. 

 For systems with higher mass-symmetry, neutrons can also be emitted in the 
comparatively longer formation stage of the CN \cite{Saxena1994} and/or from the 
acceleration phase of fission fragments \cite{Eismont1965,Hinde1984} and/or 
during neck rupture \cite{Bowman1962,Carjan2010}, which are not included in the present work.

\section{Conclusions}
   
 A statistical model description of fission / ER cross-sections and
pre-scission neutron multiplicity in heavy ion induced fusion-fission reactions
is presented by including the shell effects, the collective enhancement of
level density and the $K$-orientation effect with standard parameter values and
by treating the dissipation strength as the only adjustable parameter. It is
found that the inclusion of all the aforesaid effects eliminates the
contradictory requirements of fission hindrance for pre-scission neutron
multiplicities on one hand and fission enhancement for ER cross sections on the
other. The present work thus provides a consistent picture of fusion-fission
reactions where fission hindrance plays a role for both pre-scission emission
of neutrons and formation of evaporation residues. 

\section{Acknowledgements}

The authors thank Jhilam Sadhukhan of VECC, Kolkata for providing the form
factor of deformation dependent shell correction. One of the authors (T.B.)
acknowledges the financial support from the University Grants Commission (UGC),
Government of India.

%\section{References}

\newpage

\begin{thebibliography}{99}
\bibitem{Stephens1975} F.S. Stephens, Rev. Mod. Phys. \textbf{47}, 43 (1975).
\bibitem{Schroder1977} W.U. Schr\"{o}der and J.R. Huizenga, Ann. Rev. Nucl. Sci. \textbf{27}, 465 (1977).
\bibitem{Hinde1986} D.J. Hinde et al., Nucl. Phys. \textbf{A452}, 550 (1986).
\bibitem{Newton1988} J.O. Newton et al., Nucl. Phys. \textbf{A483}, 126 (1988).
\bibitem{BW} Niels Bohr and John Archibald Wheeler. Phys. Rev. \textbf{56}, 426 (1939).
\bibitem{Thonnensen} M. Thoennessen and G.F. Bertsch, Phys. Rev. Lett. \textbf{71}, 4303 (1993).
\bibitem{LestonePRC79} J.P. Lestone and S.G. McCalla, Phys. Rev. C \textbf{79}, 044611 (2009).
\bibitem{MahataPRC74} K. Mahata, S. Kailas, and S.S. Kapoor, Phys. Rev. C \textbf{74}, 041301(R) (2006).
\bibitem{MahataPRC92} K. Mahata et al., Phys. Rev. C \textbf{92}, 034602 (2015).
\bibitem{SagaidakPRC79} R.N. Sagaidak and A.N. Andreyev, Phys. Rev. C \textbf{79}, 054613 (2009).
\bibitem{VickyPRC89} Varinderjit Singh et al., Phys. Rev. C \textbf{89}, 024609 (2014).
\bibitem{Tatha2015} Tathagata Banerjee, S. Nath, and Santanu Pal, Phys. Rev. C \textbf{91}, 034619 (2015).
\bibitem{Tatha2016} Tathagata Banerjee et al., Phys. Rev. C \textbf{94}, 044607 (2016).
\bibitem{MayorovPRC92} D.A. Mayorov et al. Phys. Rev. C \textbf{92}, 054601 (2015).
\bibitem{YanezPRL112} R. Yanez et al., Phys. Rev. Lett. \textbf{112}, 152702 (2014).
\bibitem{SierkPRC33} A.J. Sierk, Phys. Rev. C \textbf{33}, 2039 (1986).
\bibitem{Myers1994} W.D. Myers and W.J. Swiatecki, Report No. LBL-36803, 
Lawrence Berkeley Laboratory, 1994 (unpublished).
\bibitem{MyersNP81} W.D. Myers and W.J. Swiatecki, Nucl. Phys. \textbf{81}, 1 (1966).
\bibitem{Ignatyuk1975} A.V. Ignatyuk et al., Yad. Fiz. \textbf{21}, 485 (1975) 
[Sov. J. Nucl. Phys. \textbf{21} (1975) 255].
\bibitem{Reisdorf1981} W. Reisdorf, Z. Phys. A \textbf{300}, 227 (1981).
\bibitem{BBM1974} S. Bj\o rnholm, A. Bohr and B.R. Mottelson, Proc. Int. Conf. on the Physics and Chemistry of 
Fission, Rochester 1973 (IAEA Vienna 1974) Vol. 1, p. 367.
\bibitem{JunghansNPA629} A.R. Junghans et al., Nucl. Phys. \textbf{A629}, 635 (1998).
\bibitem{PratapPRC88} Pratap Roy et al., Phys. Rev. C \textbf{88}, 031601(R) (2013).
\bibitem{ZagrebaevPRC65} V.I. Zagrebaev et al., Phys. Rev. C \textbf{65}, 014607 (2001).
\bibitem{Ohta2001}  M. Ohta, in Proceedings on Fusion Dynamics at the Extremes,
Dubna, 2000, edited by Yu. Ts. Oganessian and V.I. Zagrebaev, World Scientific, Singapore, 2001, p. 110. 
\bibitem{Ignatyuk1985} A.V. Ignatyuk, G.N. Smirenkin, M.G. Itkis, S.I. Mul'gin and 
V.N. Okolovich; Sov. J. Part. Nucl. \textbf{16(4)}, 307 (1985).
\bibitem{LestonePRC59} J.P. Lestone, Phys. Rev. C \textbf{59}, 1540 (1999).
\bibitem{FrobrichPR} P.Fr\"{o}brich and I.I. Gontchar, Phys. Rep. \textbf{292}, 131 (1998).
\bibitem{GargiPRC65} Gargi Chaudhuri and Santanu Pal, Phys. Rev. C \textbf{65}, 054612 (2002).
\bibitem{NixNPA130} J.R. Nix, Nucl. Phys. \textbf{A130}, 241 (1969).
\bibitem{Leigh1995} J.R. Leigh et al., Phys. Rev. C \textbf{52}, 3151 (1995).
\bibitem{Zebelman1973} A.M. Zebelman et al., Phys. Rev. Lett. \textbf{30}, 27 (1973).
\bibitem{Gavron1981} A. Gavron et al., Phys. Rev. C \textbf{24}, 2048 (1981).
\bibitem{Charity1986} R.J. Charity et al., Nucl. Phys. \textbf{A457}, 441 (1986).
\bibitem{Leigh1988} J.R. Leigh et al., J. Phys. G: Nucl. Phys. \textbf{14}, L55 (1988).
\bibitem{Shidling2006} P.D. Shidling et al., Phys. Rev. C \textbf{74}, 064603 (2006).
\bibitem{Bemis1986} C. E. Bemis Jr., T. C Awes, J. R. Beene, R. L. Ferguson, H. J. Kim, F. K. McGowan, 
F. E. Obenshain, F. Plasil, P. Jacobs, Z. Frankel, U. Smilansky, I. Tserruya, 
ORNL physics division progress report 6326 (1986).
\bibitem{Forster1987} J.S. Forster et al., Nucl. Phys. \textbf{A464}, 497 (1987).
\bibitem{Hinde1992} D.J. Hinde et al., Phys. Rev. C \textbf{45}, 1229 (1992).
\bibitem{Brinkmann1994} K.-T. Brinkmann et al., Phys. Rev. C \textbf{50}, 309 (1994).
\bibitem{Sikkeland1964} T. Sikkeland, Phys. Rev. \textbf{135}, B669 (1964).
\bibitem{Rossner1992} H. Rossner et al., Phys. Rev. C \textbf{45}, 719 (1992).
\bibitem{Morton1995} C.R. Morton et al., Phys. Rev. C \textbf{52}, 243 (1995).
\bibitem{Back1985} B.B. Back et al., Phys. Rev. C \textbf{32}, 195 (1985).
\bibitem{ShamlathPRC95} A. Shamlath et al., Phys. Rev. C \textbf{95}, 034610 (2017).
\bibitem{Kramers} H.A. Kramers, Physica \textbf{7}, 284 (1940).
\bibitem{JhilamPRC78} Jhilam Sadhukhan and Santanu Pal, 
Phys. Rev. C \textbf{78}, 011603(R) (2008), Phys. Rev. C \textbf{79}, 01990(E) (2009).
\bibitem{Blocki1978} J. Blocki et al., Annals of Phys. \textbf{113}, 330 (1978).
\bibitem{Wada1993} T. Wada et al., Phys. Rev. Lett. \textbf{70}, 3538 (1993).
\bibitem{PalPRC57} Santanu Pal and Tapan Mukhopadhyay, Phys. Rev. C \textbf{57}, 210 (1998).
\bibitem{HofmannPRC64} H. Hofmann et al., Phys. Rev. C \textbf{64}, 054316 (2001).
\bibitem{VickyPRC86} Varinderjit Singh et al., Phys. Rev. C \textbf{86}, 014609 (2012).
\bibitem{RohitPRC87} Rohit Sandal et al., Phys. Rev. C \textbf{87}, 014604 (2013).
\bibitem{DioszegiPRC61} I. Di\'{o}szegi, N. P. Shaw, I. Mazumdar, A. Hatzikoutelis, and P. Paul, 
Phys. Rev. C \textbf{61}, 024613 (2000).
\bibitem{HofmannPLB122}  H. Hofmann and J.R. Nix, Phys. Lett. \textbf{B122}, 117 (1983).
\bibitem{BhattPRC33} K.H. Bhatt, P. Grang\'{e}, and B. Hiller, Phys. Rev. C \textbf{33}, 954 (1986).
\bibitem{FrobrichNPA556} P. Fr\"{o}brich, I.I. Gontchar, and N.D. Mavlitov, Nucl. Phys. \textbf{A556}, 281 (1993).
\bibitem{JhilamPRC81} Jhilam Sadhukhan and Santanu Pal, Phys. Rev. C \textbf{81}, 031602(R) (2010).
\bibitem{NadtochyPRC85} P.N. Nadtochy et al., Phys. Rev. C \textbf{85}, 064619 (2012).
\bibitem{Vanin1999} D.V. Vanin, G.I. Kosenko, and G.D. Adeev, 
Phys. Rev. C \textbf{59}, 2114 (1999).
\bibitem{Goodman1986} Alan L. Goodman, Phys. Rev. C \textbf{34}, 1942 (1986).
\bibitem{Saxena1994} A. Saxena, A. Chatterjee, R. K. Choudhury, S. S. Kapoor, and 
D. M. Nadkarni, Phys. Rev. C \textbf{49}, 932 (1994).
\bibitem{Eismont1965} V. P. Eismont, Sov. J. At. Energy \textbf{19}, 1000 (1965).
\bibitem{Hinde1984} D. J. Hinde, R. J. Charity, G. S. Foote, J. R. Leigh, J. O. Newton, S. Ogaza, 
and A. Chatterjee, Phys. Rev. Lett. \textbf{52}, 986 (1984).
\bibitem{Bowman1962} Harry R. Bowman, Stanley G. Thompson, J. C. D. Milton, and Wladyslaw J. Swiatecki, 
Phys. Rev. \textbf{126}, 2120 (1962).
\bibitem{Carjan2010} N. Carjan and M. Rizea, Phys. Rev. C \textbf{82}, 014617 (2010).
\end{thebibliography}
\end{document}